\documentclass[aps,pra,reprint,nofootinbib,showpacs]{revtex4-1}
\usepackage{amsfonts}
\usepackage{amsmath,amssymb}
\usepackage{braket}
\usepackage{tikz}
\usepackage{booktabs}
\usepackage{graphicx}
\usepackage{blkarray}
\usepackage{float}
\usepackage{multirow}
\usepackage{bm}
\usepackage{xcolor}
\usepackage{hyperref}
\hypersetup{
     colorlinks   = true,
     citecolor    = blue
}

\def\be{\begin{equation}}
\def\ee{\end{equation}}
\def\bea{\begin{eqnarray}}
\def\eea{\end{eqnarray}}

\newcommand{\diagonal}{\mathrel{\rotatebox[origin=c]{45}{$\leftrightarrow$}}}
\newcommand{\antidiagonal}{\mathrel{\rotatebox[origin=c]{135}{$\leftrightarrow$}}}
\newcommand{\horizontal}{\ensuremath{\leftrightarrow}}
\newcommand{\vertical}{\ensuremath{\updownarrow}}

\newcommand{\sml}[1]{\scalebox{0.85}{$ #1 $} }

\DeclareMathSizes{10}{10}{8}{5}

\definecolor{mygreen}{RGB}{1,128,6}
\definecolor{mymgnta}{RGB}{186,0,193}

\usepackage{makecell}
\setcellgapes{7pt}
\usepackage[caption=false]{subfig}

\begin{document}
	\title{Experimental decoy-state asymmetric measurement-device-independent quantum key distribution over a turbulent high-loss channel}
	\author{Kazi Reaz}
	\email{kreaz@vols.utk.edu}
	\affiliation{Department of Physics and Astronomy, The University of Tennessee, Knoxville, TN 37996-1200, USA}	

\author{Md Mehdi Hassan}
\email{mhassa11@vols.utk.edu}
	\affiliation{Department of Physics and Astronomy, The University of Tennessee, Knoxville, TN 37996-1200, USA}
 
 \author{Adrien Green}
	\email{agreen91@vols.utk.edu}
	\affiliation{Department of Physics and Astronomy, The University of Tennessee, Knoxville, TN 37996-1200, USA}

	\author{Noah Crum}
	\email{ncrum@vols.utk.edu}
	\affiliation{Department of Physics and Astronomy, The University of Tennessee, Knoxville, TN 37996-1200, USA}

	\author{George Siopsis}
	\email{siopsis@tennessee.edu}
	\affiliation{Department of Physics and Astronomy, The University of Tennessee, Knoxville, TN 37996-1200, USA}
	\date{\today}
	\begin{abstract}
        Real-world BB84 Quantum Key Distribution (QKD) systems utilize imperfect devices that introduce vulnerabilities to their security, known as side-channel attacks. Measurement-Device-Independent (MDI) QKD authorizes an untrusted third party to make measurements and removes all side-channel attacks. The typical implementations of MDI-QKD employ near symmetric channels which are difficult to realize physically in many practical scenarios such as when asymmetric channel losses are present, normally a consequence of the communication environment. Maritime and satellite-based communications are two such instances in which the channels are characterized by continuously changing losses in different channels. In this work, we perform asymmetric MDI-QKD in a laboratory environment with simulated turbulence using an Acousto-Optic Modulator (AOM) to interrogate the performance of free-space quantum communication. Under turbulent conditions, scattering and beam wandering cause intensity fluctuations which decrease the detected signal-to-noise ratio.  Using the 7-intensity optimization method proposed by Wang et al., coupled with Prefixed-Threshold Real-time Selection (P-RTS), we demonstrate enhancement in the secure key rate under turbulent conditions for finite-size decoy-state MDI QKD. Furthermore, we show that P-RTS can yield considerably higher secure key rates for a wide range of atmospheric channel parameters.
	\end{abstract}
	\maketitle

\section{Introduction}
Although QKD has been proven to be unconditionally secure theoretically, practical systems have back doors that Eve can exploit due to device imperfections. In particular, detectors can be attacked through various approaches, such as the Blinding the Detector attack \cite{lydersen2010}, Phase-Remapping attack \cite{chi-hang2007}, time-shift attack \cite{bing-qi2007}, as well as through other means (see Jain, et al.\ \cite{nitin2016}). Under these considerations, Lo, Curty, Qi \cite{lo2012} proposed the MDI-QKD protocol which removes the need for detector security under the condition that Alice and Bob can prepare near-perfect quantum states. So far, implementations of MDI-QKD have been performed in nearly symmetric channels \cite{silva2013, rubenok2013, yang2013, tang2014, yanlin2014, hualei2016, xu2013}, however
symmetric channels are difficult to realize in practical scenarios. For example, in a free-space implementation, Alice's and Bob’s channels have different losses due to being in geographically different locations. One proposal to balance this asymmetry is to add extra loss in one channel through the addition of extra fiber, which, however, lowers the key rate \cite{xu2013}. Moreover, an MDI-QKD implementation in a maritime environment between ships or a satellite-based system will experience continuously changing losses in the different channels that cannot be removed with additional fiber. To overcome these issues, the authors of Refs.\ \cite{xu2013, wang2013} proposed that asymmetric decoy state intensities  be used to generate a higher key rate instead of adding fiber to one of the channels. Wang, Xu, and Lo \cite{wang2019} provided theoretical optimizations for 7 different decoy state intensities that have given the highest expected secure key rates thus far in the literature.

For our experiment, we conducted simulations to replicate atmospheric effects on traveling pulses, aiming to mimic real-world scenarios. When a signal moves through fiber, attenuation occurs due to various reasons (e.g., absorption, scattering) but the loss remains relatively uniform in time. In contrast, a free-space channel suffers variable attenuation as a result of weather (temperature, clouds, dust, etc.) and altitude.  Fortunately, through consideration of the signal's wavelength and the presence of turbulence, there exist well-established models for statistically describing the free-space optical channel.

In 2012, Erven, et al.\ \cite {erven2012} proposed a signal-to-noise-ratio filter (SNRF) in the post-processing stage to increase the key rate. After data collection, the bits are arranged into time blocks whose duration is adaptive, depending on the detection rate. By optimizing the block duration, an optimum threshold was achieved. In this protocol, the channel loss was assumed to be static (equal to the mean loss), which may not hold under conditions of strong turbulence and in the high-loss regime.

In 2015, Vallone, et al.\ \cite{vallone2015} employed an auxiliary classical laser beam that co-propagates alongside the quantum channel. The classical beam exhibits loss proportionally to the quantum channel and therefore can be used to post-select high transmittance quantum data to reduce the average error rate. This introduces an optimization problem, as discarding signals from poor transmittance periods can reduce the quantum bit error rate (QBER), yet discarding too many can eventually decrease the secure key rate. Finding the optimal threshold transmittance in the log-normal distribution is therefore critical to maximize the key rate. 

A protocol that utilizes a pre-fixed threshold (P-ARTS) was introduced theoretically in 2018 \cite{wang2018}, and subsequently demonstrated experimentally across different channel losses \cite{lefty2021,mehdi2023} in the context of finite key decoy state BB84. It was demonstrated in \cite{wang2018} that the optimal threshold is only dependent on the transmittance if the device parameters (e.g., detector efficiency, dark count, source intensities) remain fixed. Since this threshold can be predetermined, it facilitates real-time data filtering, resulting in savings in storage memory and analysis time.

In our experiment, we implemented asymmetric MDI QKD following \cite{wang2019}. We used 7 asymmetric intensities and the decoupled bases method in a protocol for asymmetric channels. We analyzed the P-ARTS method in this context using a signal wavelength of 1550 nm and an average channel loss between 30 and 33 dB with moderate turbulence, which we modeled as a log-normal distribution. We tested the theoretical assumptions of the P-RTS theory in this context and found significant improvements in the key rate compared to using no data rejection, especially for high loss.

The structure of this paper is as follows: In Section \ref{sec:2}, we discuss the protocol of polarization-encoded MDI QKD including decoy states, and atmospheric turbulence and channel loss implementation. Our experimental setup and all relevant parameters are detailed in Section \ref{sec:3}. Experimental data are presented and compared with simulations in Section \ref{sec:4}. Finally, in Section \ref{sec:5} we present our conclusions. All essential equations for our calculations can be found in Appendix \ref{sec:A1}. Appendix \ref{sec:A2} details the classical channel we employed.        


\section{Theory} \label{sec:2}

Here we describe the theory underpinning MDI-QKD and expound upon the instance of asymmetric channels. Specifically, we outline the polarization encoding scheme and describe the implementation of asymmetric MDI-QKD using the 7-intensity method introduced in \cite{wang2019}. We also outline the atmospheric model under consideration for severe channel loss with a moderate level of turbulence. 

\subsection{Asymmetric MDI QKD}
\begin{figure}[ht!]
    \centering
    \includegraphics[width=0.48\textwidth]{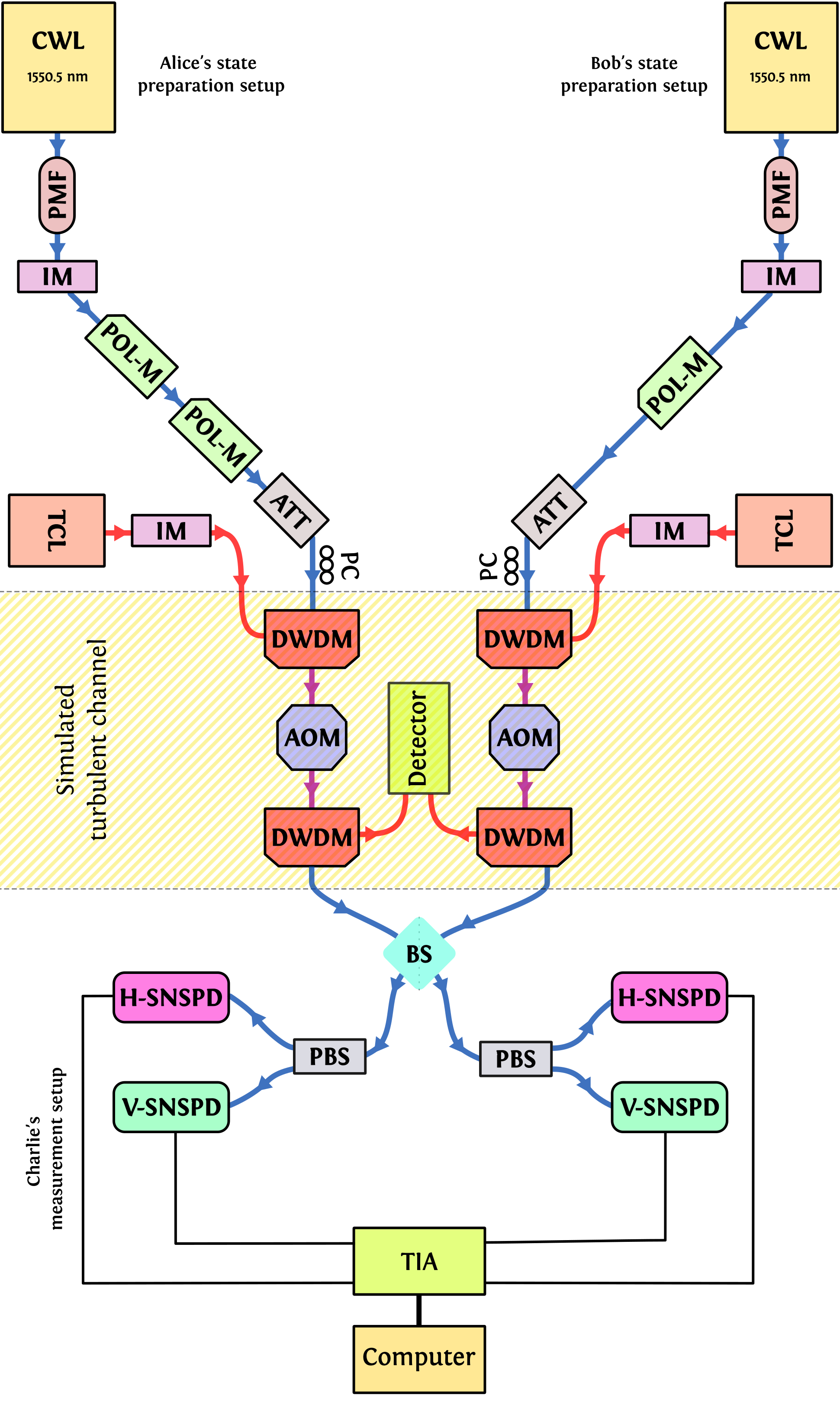}
    \caption{ Schematic of the experimental setup for the asymmetric MDI QKD system: Alice (left) and Bob (right) generate randomly polarized signal pulse trains using a continuous-wave laser (CWL), polarization-maintaining fiber (PMF), intensity modulator (IM), polarization modulator (POL-M), and attenuator (ATT). A tunable continuous-wave laser (TCL) with an intensity modulator (IM) is used to create classical probe pulses. The simulated turbulent channel consists of dense wave division multiplexers (DWDM) and acousto-optic modulators (AOM). Charlie measures the incoming signal using a beam splitter (BS), polarization beam splitters (PBS), superconducting nanowire single-photon detectors (SNSPDs), and a time interval analyzer (TIA) connected to a computer.}
    \label{fig:f_setup}
\end{figure}

MDI-QKD is designed to automatically remove all detector side-channels by employing time-reversed entanglement. In this protocol, Alice and Bob send light pulses to a third party, Charlie, who possesses a Bell-state analyzer based on linear optics and single-photon detection. Charlie projects the input photons to Bell states and publicly announces the measurement results, which allows Alice and Bob to generate a secret key after classical post-processing. Alice and Bob may choose time-bin encoding \cite{yang2013, kaneda2017}, phase encoding \cite{pirandola2015}, or polarization encoding \cite{ma2012, silva2013, tang2014}. In this work, we use polarization encoding. The Bell-state analyzer in MDI-QKD relies on the Hong-Ou-Mandel (HOM) effect \cite{hong1987} where photons from Alice and Bob interfere at a 50:50 beam splitter. A high HOM visibility can usually be translated into a low QBER and therefore a high secret key rate. To achieve a high HOM visibility, photons from Alice and Bob should be indistinguishable in all degrees of freedom. Furthermore, when MDI QKD is implemented with weak coherent sources, a high HOM visibility requires the average photon numbers from Alice and Bob to be matched at the beam splitter \cite{tang2014}. With polarization encoding, Alice and Bob encode their random bits on the polarization of their respective weak coherent states, using one of two bases, rectilinear (Z) or diagonal (X), and Charlie performs Bell-state measurements using a setup depicted in Fig. \ref{fig:f_setup}. A bit of raw key is generated whenever Charlie measures a coincidence of photons with orthogonal polarizations (D$_1$H, D$_1$V, D$_2$H \& D$_2$V) using a set of four single-photon detectors, and Alice and Bob use the same encoding basis. Since photons are bosons with integer spin, we can write their overall state as $\ket{\Psi} = \ket{\Psi_{\text{spatial}}} \otimes \ket{\Psi_{\text{polarization}}}$. Due to the HOM effect, if the photons come out of opposite sides of the beam splitter, both $\ket{\Psi_{\text{spatial}}}$ and $\ket{\Psi_{\text{polarization}}}$ must be antisymmetric and the polarization state should be $\ket{\Psi^-} = \frac{1}{\sqrt{2}} \left(\ket{HV} - \ket{VH}\right)$. If the photons come out of the same port, then both must be symmetric, so the polarization state should be $\ket{\Psi^+} = \frac{1}{\sqrt{2}} \left(\ket{HV} + \ket{VH} \right)$. The other two possible polarization states,  $\frac{1}{\sqrt{2}} \left(\ket{HH} \pm \ket{VV}\right)$, are not identifiable with our detector setup. When Alice and Bob receive the measurement result from Charlie, they can easily determine the bits they sent. If Charlie announces $\ket{\Psi^+}$  or $\ket{\Psi^-}$ and both Alice and Bob used the rectilinear basis, then one of them has to perform a bit flip to his/her bit. If both used the diagonal basis and Charlie announces $\ket{\Psi^+}$, then no bit flip is necessary, but if $\ket{\Psi^-}$ is announced, one of them must perform a bit flip. MDI-QKD is free from any attack on the detectors, but it is not immune to attacks on sources. So, our phase-randomized weak coherent pulses must be protected from the photon number splitting attack. In our lab, we used CW laser sources which have a non-zero probability of multiple photon pulses. To prevent the photon number splitting attack, decoy states have been implemented \cite{xiangbin2005, hwang2003, lo2005}. 


In the two-user instance of asymmetric MDI-QKD, Alice and Bob utilize quantum channels with asymmetric transmittances $\eta_{A}$ and $\eta_{B}$, resepctively, where $\eta_{A} \neq \eta_{B}$. They must choose optical intensities $s_A$ and $s_B$, respectively, such that the resulting key rate is maximal \cite{xu2013}. The typical choice is to select intensities obeying $s_A \eta_A = s_B \eta_B$, which ensures a symmetry of photon flux at the relay position, Charlie, providing higher-quality HOM interference \cite{hong1987}. This approach is sub-optimal in the asymmetric setting and can even result in zero key rate for highly asymmetric channels. In particular, HOM interference is dependent on errors only in the X basis, namely the phase error rate, and not those in the Z basis, the bit error rate. An optimal approach to key generation requires decoupling the decoy state estimation performed in the X basis from that of the bit generation in the Z basis \cite{wang2019}. 

In the 7-intensity optimization method of \cite{wang2019}, Alice and Bob select a set of four intensities each. These intensities correspond to the signal state intensities $\{s_A, \,s_B\}$ in the Z basis, and the decoy state intensities in the X basis, $\{\mu_A, \,\nu_A, \,\omega\}$ and $\{\mu_B, \,\nu_B, \,\omega\}$, for Alice and Bob, respectively. These choices constitute seven separate intensities each paired with the probability of their preparation. As indicated above, the X basis is reserved for decoy state analysis while the Z basis is used to establish the secret key. Therefore, the X-basis intensities are selected to ensure high HOM visibility at the central relay by compensating for the channel asymmetry. This selection provides symmetry of the photon-flux at Charlie and roughly satisfies $\frac{\mu_A}{\mu_B} = \frac{\nu_A}{\nu_B} \approx \frac{\eta_B}{\eta_A}$. Due to the decoupling of bases, the signal-state intensity is a free parameter and can be adjusted independently to provide an optimal key rate. In general, this approach does not satisfy $\frac{s_A}{s_B} = \frac{\eta_B}{\eta_B}$. Altogether, Alice and Bob have a set of 12 parameters to optimize, their intensities and the associated probabilities of preparation; namely, $ \{s_{A}, \,\mu_{A}, \,\nu_{A}, \,p_{s_{A}}, \,p_{\mu_{A}}, \,p_{\nu_{A}}, \,s_{B}, \,\mu_{B}, \,\nu_{B}, \,p_{s_{B}}, \,p_{\mu_{B}}, \,p_{\nu_{B}}\} $.



\subsection{Simulating a Turbulent Atmosphere}

In our experiment, we chose the average channel loss at 30-33 dB which is considered as severe channel loss with a moderate level of turbulence. Similarly to our previous work \cite{lefty2021, mehdi2023}, we chose the standard and well accepted log-normal distribution to model the probability distribution of the transmittance coefficient (PDTC). Mathematically, 
\begin{equation}
\scalebox{0.95}{$p_{_{\eta_o, \sigma}}(\eta) = \dfrac{1}{\sqrt{2\pi} \sigma \eta } \exp \left\{ -\frac{\big[\ln{(\frac{\eta}{\eta_o})} + \frac{\sigma^2}{2} \big]^2}{2\sigma^2} \right\}  $} \label{eqn:1}
\end{equation}
It depends on two parameters, namely $\eta_o$ (average channel loss) and $\sigma^2$ (logarithmic
irradiance variance). The latter, commonly known as Rytov parameter, is correlated with turbulence. If the wavelength remains stable throughout the implementation of the protocol, the plane-wave approximation yields: $\sigma^2 = 1.23 C_n^2  k^{\frac{7}{6}} L^{\frac{11}{6}}$, 
where $k$ is the wave number, $C_n^2$ is the refractive index structure parameter ($n$ being the refractive index), and $L$ is the distance traveled by the wave. While, typically, $C_n^2$ is an intricate function influenced by factors such as time of day, local wind conditions, solar elevation angle, and terrain type, most scenarios can be adequately addressed with a simple mathematical relation connecting $C_n^2$ and altitude.


\section{Experimental Setup} \label{sec:3}

Our experimental setup is sketched in Figure \ref{fig:f_setup}.

To create signals, Alice and Bob use identical continuous lasers (CWL) as their sources. The semiconductor based CW lasers are mode locked at central wavelength 1550.5 nm in 2 mW low power output. Polarization maintaining fiber (PMF) carries the beam into an intensity modulator. The conversion from continuous wave to pulse train is carried out by LiNbO$_3$ intensity modulator. The intensity modulators (IM) are driven by an arbitrary waveform generator (Tektronix) to control the intensity level of each pulse in order to implement the desired signal and decoy states with different mean photon numbers. In our case, the full-width half maximum (FWHM) pulses were $\sim \!1$ ns at a 10 MHz repetition rate. DC bias voltages were precisely controlled by a null point modulator bias controller device to achieve high extinction ratio by applying compensation bias voltage.

In the next stage, Alice and Bob encode the desired polarization state into the pulses using Polarization Modulators (PM). Each PM consists of a polarization controller, a beam circulator, a phase modulator, and a Faraday mirror. The phase modulator is driven by an arbitrary wave function generator. Alice's side contains an extra polarization modulator to align diagonal and anti-diagonal polarization \cite{tang2016}. To reach the single photon level, attenuation was applied to the pulses. In our experiment, we used both digital and manual attenuators.  
The pulses were attenuated to the single photon level with the help of variable attenuators (ATT). Polarization controllers (PC) were utilized by both parties to fine tune the polarization states to ensure good HOM visibility while calibrating the setup before recording the data. 

Next, the quantum states are multiplexed with a classical beam with the help of a 200 GHz DWDM (dense wavelength-division multiplexing) device. 
For the classical signal, we used tunable classical lasers (TCL), tuned at 1554 nm. Another set of  IMs was used to convert the continuous beam into a pulse train with a pulse rate at 4-kHz repetition at 3 ns FWHM. Classical and quantum signals were muxed in a DWDM device in ITU channel 29 and 33.5, respectively. The classical pulses were used to probe the channel's transmittance statistics. The detailed explanation of estimating the channel's transmittance with classical probe pulses is shown in Appendix \ref{sec:A2}. 
The mixed signals (quantum and classical) are directed into an AOM (Acousto-Optic Modulator) device independently in each side, which are used to simulate the desired atmospheric channel loss model. In our experiment, Alice's channel suffers a different channel loss as compared to Bob's channel. Another set of DWDM devices were used to filter out (demux) the classical signal from the quantum signal. The classical signal was then detected by a high-gain detector and analyzed with an oscilloscope.

Quantum signals from both sides are fed into a 50:50 beam splitter (BS). Since the setup is properly calibrated, upon interaction, the photons emerge through the output terminal(s) and each basis is resolved by a polarization beam splitter. The outputs of each terminal are detected by Superconducting Nano-wire Single Photon Detectors  (SNSPD). All the detections of our SNSPDs were recorded by a Time Interval Analyzer (TIA) from IDQ. The TIA was connected to a computer to analyze the count rate of each channel and the coincidence detections among them. 

\begin{table}[ht!]
\makegapedcells
\caption{\raggedright \footnotesize List of Optimized Parameters}
\begin{tabular}{ |c@{\hspace{5mm}}  | c |  c | c | c | c | c | c | c |  }
\specialrule{2pt}{0pt}{2pt}\hline
 & Channel & Loss & $s$ & $\mu$ & $\nu$ & $p_{s}$ & $p_{\mu}$ & $p_{\nu}$ \\\hline
\multicolumn{1}{|c|}{\multirow{2}{*}{\rotatebox{90}{\textbf{30 dB}}}} & Alice & 25 dB & 0.593 & 0.427 & 0.101 & 0.588 & 0.045 & 0.238 \\\cline{2-9}
\multicolumn{1}{|c|}{} & Bob & 5 dB & 0.181 & 0.076 & 0.018 & 0.601 & 0.041 & 0.240 \\\specialrule{2pt}{0pt}{0pt}
\multicolumn{1}{|c|}{\multirow{2}{*}{\rotatebox{90}{\textbf{33 dB}}}} & Alice & 25 dB & 0.556 & 0.463 & 0.114 & 0.597 & 0.039 & 0.245 \\\cline{2-9}
\multicolumn{1}{|c|}{}& Bob & 8 dB & 0.192 & 0.088 & 0.021 & 0.582 & 0.035 & 0.249  \\\hline
\end{tabular}
\label{table:1}
\end{table}

Before conducting the experiment, we calculated the optimized intensities. The signal state intensities in the Z basis (s), decoy state intensities in the X basis \sml{(\mu, \,\nu, \,\omega)}, and the associated probabilities \sml{( p_s, \,p_\nu, \,p_\nu )}. We optimize our decoy parameters and probabilities stochastically using a genetic algorithm which is a preferred technique because it does not require any initial condition. The optimized signal and decoy parameters are given in table \ref{table:1}.

For our experiment, we sent $10^{12}$ pulses and collected data for 27 hours at a 10-MHz repetition rate.  Prior to the experiment, we chose the total number of pulses (N), the Z basis misalignment $e_d^Z$, the X basis misalignment $e_d^X$, the expected dark counts $Y_0$, the detector efficiency $\eta_D$, and an estimated channel transmittance $\eta_0$, as shown in table \ref{table:2}. We chose the Rytov parameter value to be 1 (moderate turbulence). The noise model and finite key calculation for our experiment are explained in Appendix \ref{sec:A1}.

\begin{table}[ht!]
\caption{\raggedright \footnotesize List of Experimental Parameters}
\begin{tabular}{l@{\hspace{12mm}} r@{\hspace{2mm}}}
\specialrule{2pt}{0pt}{1pt}\hline\addlinespace[0.5ex]
\textbf{Parameter} & \textbf{Value} \\ \midrule\addlinespace[1ex]
Number of pulses (N) & 1$\times10^{12}$ \\
Detector efficiency & 0.84$\pm$0.04 \\
Dead time & $\sim$80 ns  \\
Charlie's optical efficiency &  0.42$\pm$0.02 \\
Time jitter &  $\leq 50$ ps \\
\multicolumn{2}{c}{\multirow{2}{*}{ \textbf{Polarization Error}}}\\\addlinespace[2.5ex]
Rectilinear basis ($e_{\text{dz}}$) & 0.004$\pm$0.002\\
Diagonal basis ($e_{\text{dx}}$) & 0.02\\
\multicolumn{2}{c}{\multirow{2}{*}{\textbf{Detector Dark Count Probability}}}\\\\
\textbf{Detector} & \textbf{Probability($\times 10^{-7}$)} \\ \addlinespace[0.5ex]
Horizontal  ($Y_{0}^{\horizontal}$) & $4.1 \pm 0.6$  \\
Vertical  ($Y_{0}^{\vertical}$) & $3.7 \pm 0.6$  \\
Diagonal  ($Y_{0}^{\tiny{\diagonal}}$) & $3.2 \pm 0.6 $\\
Antidiagonal  ($Y_{0}^{\tiny{\antidiagonal}}$) & $ 3.6 \pm 0.6 $ \\ \addlinespace[0.5ex]
\specialrule{2pt}{0pt}{1pt}
\end{tabular}
\label{table:2}
\end{table}

\section{Analysis}
\label{sec:4}

After extracting the sets $n_{i,j}^{X}, \, m_{i,j}^{X}$, where $i, j = \mu, \nu, \omega$ (see Appendix \ref{sec:A1} for more details), we calculated the secure key rate using Eq.\ \eqref{eq:A18},
\begin{align}
 R = P_{s_{A}}P_{s_{B}} & \Big\{(se^{-s})^{2} \; Y_{11}^{X,L} \; \left[ 1-h_{2}(e_{11}^{X,U}) \right] \nonumber\\
 &- f \; Q_{ss}^{Z} \; h_{2} \; (E_{ss}^{Z}) \Big\}   \label{eqn:2}
\end{align}
where $Y_{11}^{X,L}$ is the lower bound of the single photon yield and $e_{11}^{X,U}$ is the upper bound of the single photon QBER in the X-basis estimated from decoy-state statistics. $Q_{ss}^{z}$ and $E_{ss}^{Z}$ are the gain and QBER, respectively, in the Z-basis which can be determined from experimental data directly. $f$ quantifies the error correction efficiency and $h_{2}$ is the binary entropy function, given by 
$  h_{2}(x)=-x \log_{2}(x) - (1-x) \log_{2}(1-x) $.

\begin{figure}[t]
    \centering
    \includegraphics[width=0.48\textwidth]{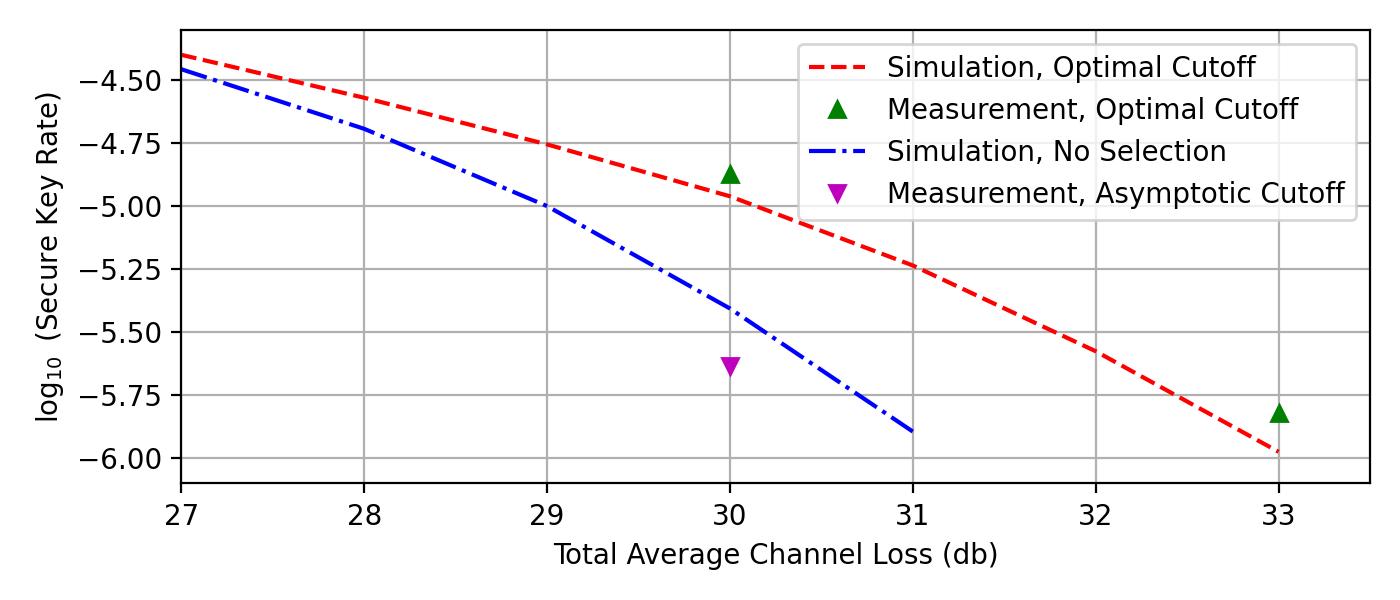}
    \caption{ Using the P-RTS method, simulations (dashed lines) and measurements (arrowheads) were conducted. The secure key rate improves when an optimized cutoff is applied, as compared to the asymptotic cutoff in both simulation and measurement. The asymptotic cutoff does not generate any practical key rate at 33 dB in either case. }
    \label{fig:f_simulation}
\end{figure}

\begin{figure}[ht]
\includegraphics[width = .9\linewidth]{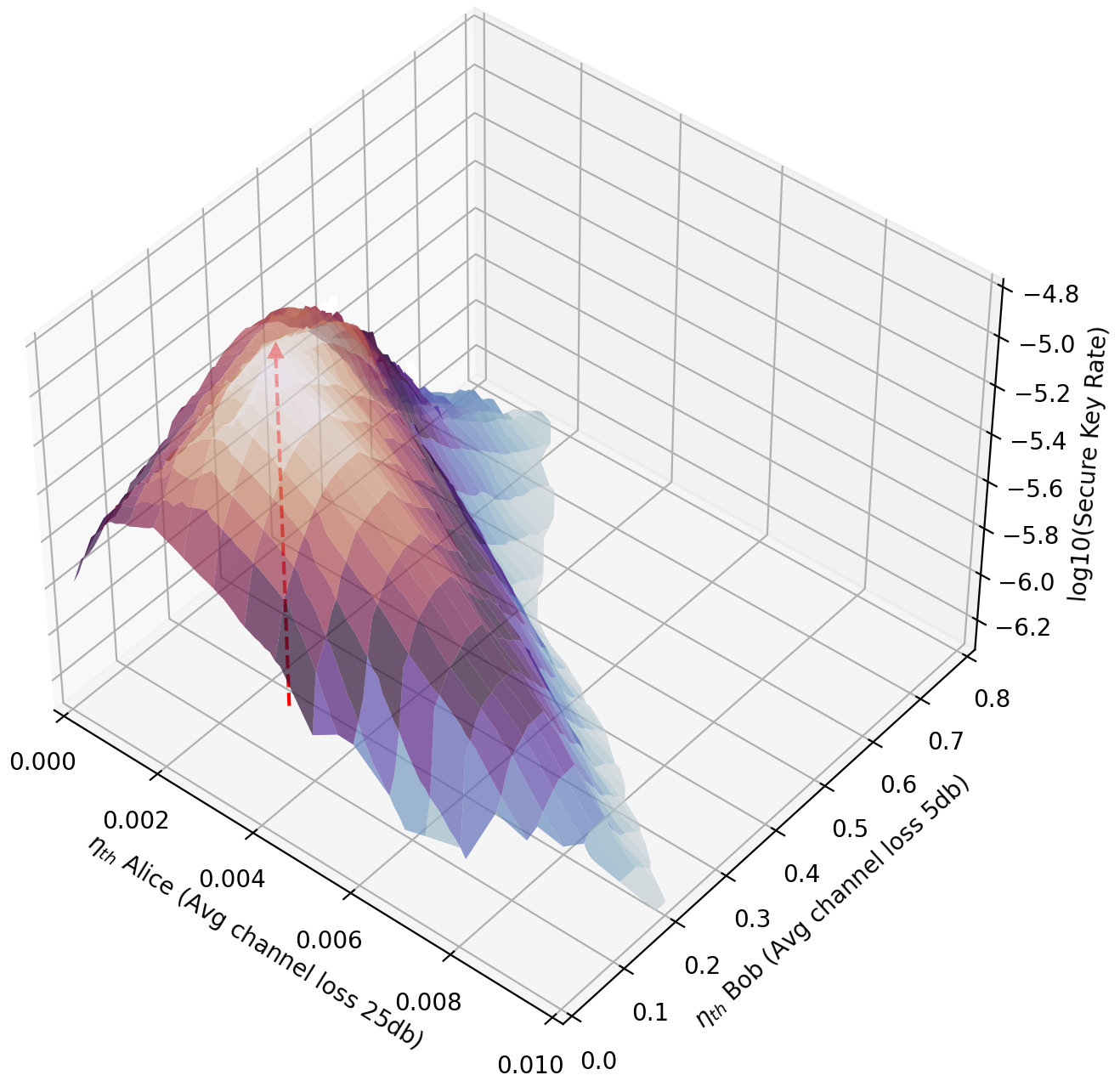}    
\includegraphics[width= .9\linewidth]{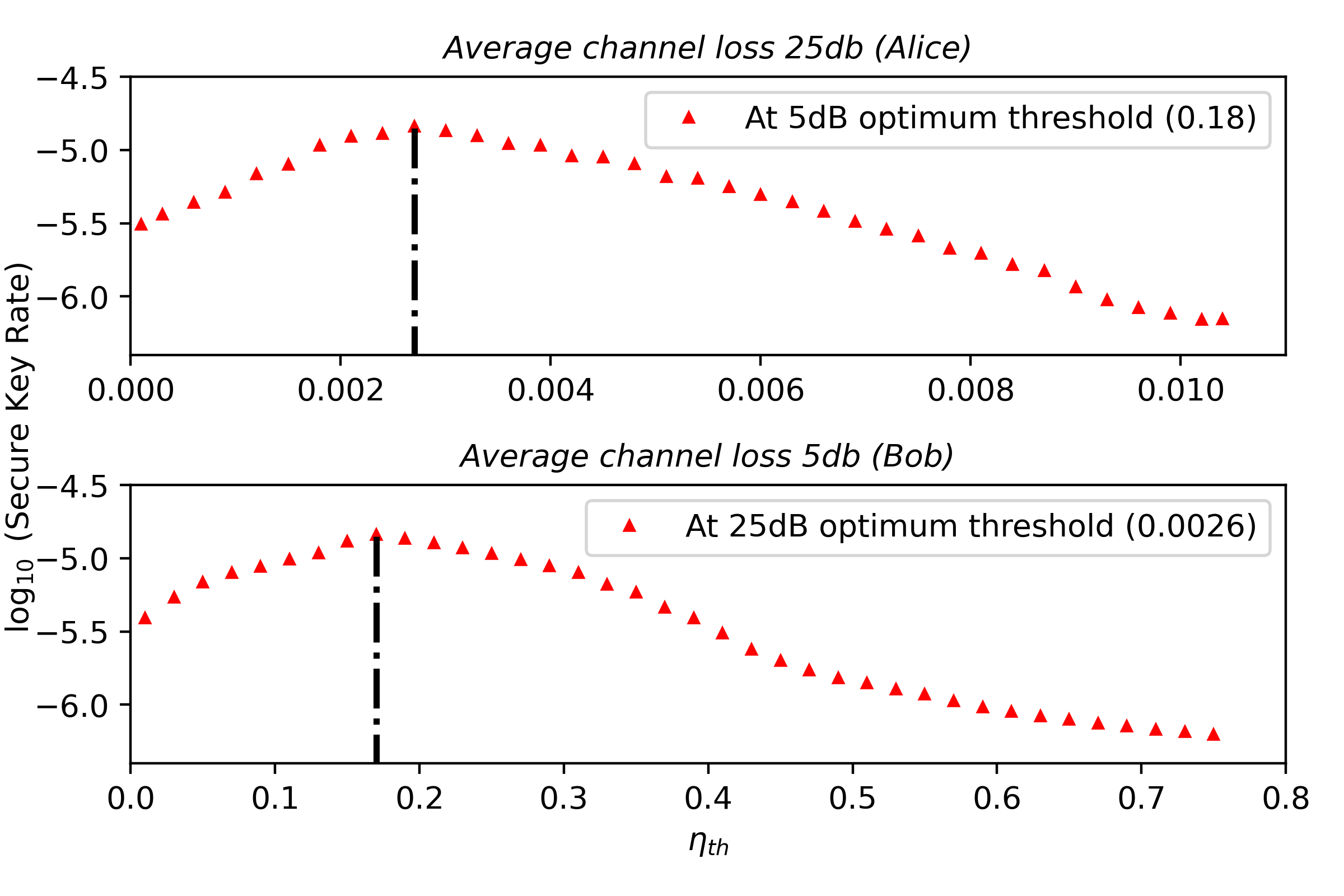} 
\caption{Finding optimal threshold point for asymmetric MDI QKD at total 30 dB mean channel loss using ARTS type distribution. Upper panel: 3D plot of all the measurement data points correspond to ARTS type post selection where we scanned successive transmittance cutoffs $\eta_{\text{th}}$ and extracted the corresponding $\log $ of secure key rate $R$  at 30 dB total average channel loss. For Alice, $\eta_{\text{th}} = 0.0026$, and  for Bob, $\eta_{\text{th}} = 0.18$. Lower panel: An example of the cross-section of the 3D plot at the optimal cut off (red arrow) point.} \label{fig:3} %
\end{figure}

The main objective of our experiment was to apply P-RTS prior to data collection to find a threshold point. A comparison was also made to the brute force ARTS method, where all sets of transmittance cut-offs were checked to achieve the highest secure key rate. ARTS always gives the highest secure key rate, but we wanted to show that by using P-RTS we could achieve the same results. In this process, knowledge of the whole distribution was not required. 

We calculated the secure key rate as a function of mean channel loss in different ways. We simulated the secure key rate vs $\eta_{\text{th}}$ for both zero cut-offs and for optimal cut-off conditions. Fig.\ \ref{fig:f_simulation} shows the asymptotic threshold cutoff (static) and optimized cut-offs over the examined mean channel loss. The graph shows that P-RTS gives a substantially higher key rate compared to the static case, and in particular, allows for key to be generated with $\sim \! 2$ dB additional loss. Our experimental secure key rate is very close to the simulation curve in both cases. The reasons for the small deviation are optical misalignment and fluctuation in the average signal and decoy photon number and the difference between measured detector efficiencies. There is an uncertainty of $\pm 0.005$ in our setting during the experiment for the desired signal photon number ($s_{A}$ and $s_{B}$) and weak decoy photon number  \scalebox{0.9}{$(\mu_{A}, \,\nu_{A}, \,\omega_{A}, \,\mu_{B}, \,\nu_{B}, \,\omega_{B})$}  given in Table \ref{table:1}. 

We present our measurement results in Fig. \ref{fig:3} (upper panel) for total mean channel loss of 30 dB, with Alice's channel experiencing an average channel loss of 25 dB and Bob's an average of 5 dB loss. At this level of asymmetry between the channels, the near-symmetric technique does not generate a secure key \cite{huiliu2019}. To derive the optimized threshold, we studied the distilled key rate as a function of threshold transmittance $\eta_{\text{th}}$ for both channels. Here, all the measurement data points correspond to ARTS \cite{gisin2006} type post-selection, where we scanned successive transmittance cut-offs and extracted the corresponding secure key rate. The red arrow corresponds to the optimal cut-off. We observed optimal threshold at $\eta_{\text{th}} =0.0026$ for average channel loss of 25 dB, and $\eta_{\text{th}} =0.18$ for average channel loss of 5 dB. 

Fig.\ \ref{fig:3} (lower panel) shows an example of the cross-section of the 3D plot at optimal cutoff (red arrow). These cross-sections are shown to illustrate the optimal cut off selection point.

\begin{figure}[ht]
    \centering
    \includegraphics[width=0.9\linewidth]{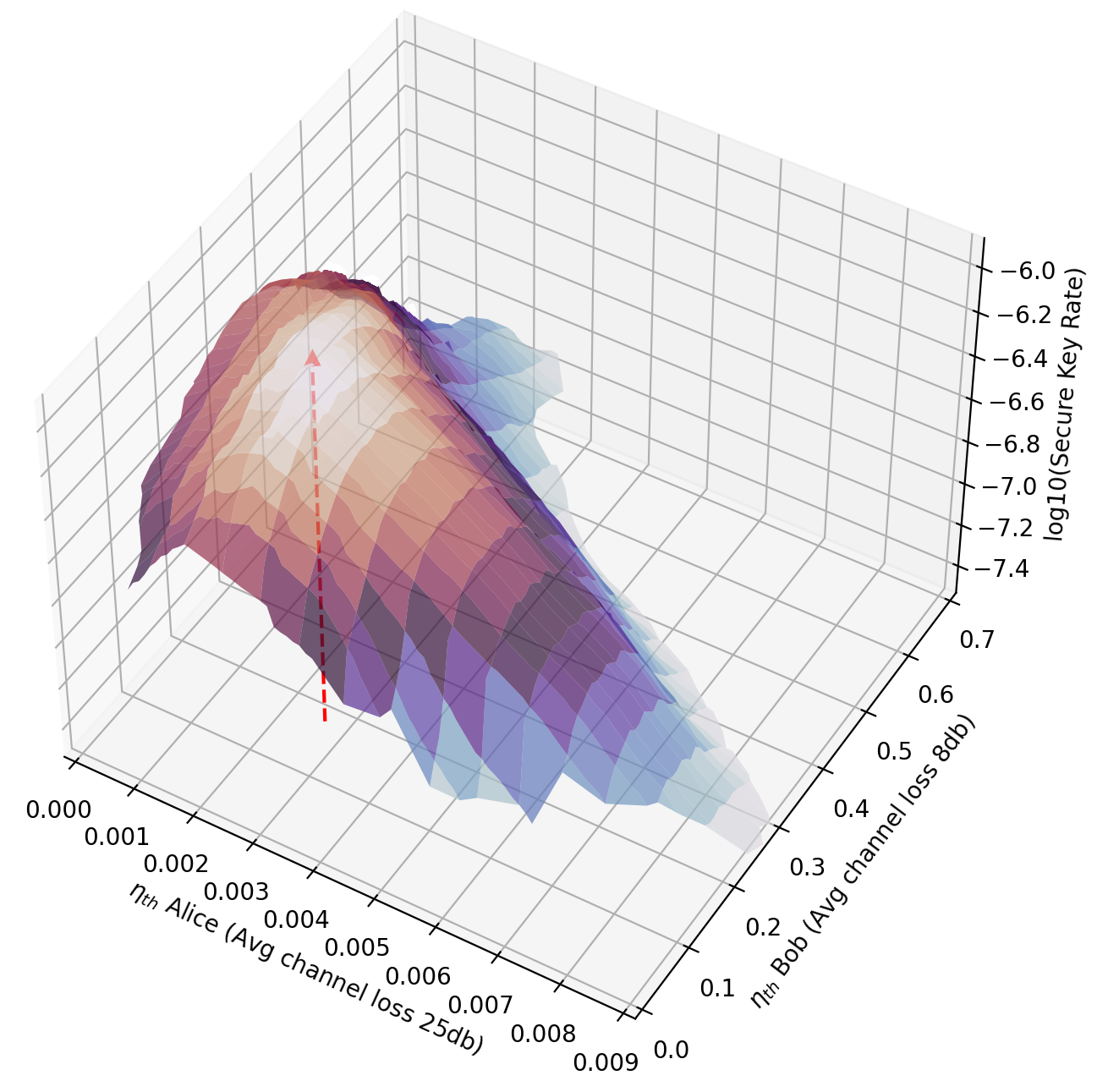}
    \caption{ 3D plot of ARTS type measurements: $\log $ of the secure key rate $R$ for increasing applied transmittance cutoff $\eta_{\text{th}}$  at 33 dB total average channel loss where Alice experiences 25 dB mean channel loss and Bob experiences 8 dB mean channel loss. }
    \label{fig:f_skr33}
\end{figure}

Fig. \ref{fig:f_skr33} shows ARTS-type post-selection data points for a total of 33 dB average channel loss (Alice at 25 dB and Bob at 8 dB). Here, the asymptotic threshold cut-off point does not generate a finite key rate. In contrast, applying P-RTS provides a good key rate which closely matches our simulation. The red arrow in Fig.\ \ref{fig:f_skr33} points to the optimal cutoff.






\section{Conclusion}
\label{sec:5}

In conclusion, we implemented an experimental demonstration of decoy-state MDI-QKD with asymmetric channels in a free-space environment where Alice and Bob are at different distances from Charlie and whose channels experience different channel losses. Our study used a realistic log-normal model to describe moderate atmospheric turbulence.

Our experiment showed that the P-RTS method finds the same optimal cutoff that would be found using the ARTS method, demonstrating proof of the P-RTS theory in the context of asymmetric, finite-key decoy-state MDI QKD and the secure key rate can be significantly improved in turbulent atmospheric conditions, especially at high loss. This selection method can be seamlessly integrated without major technological upgrades, saving computational resources. 

It is worth noting that one could combine both P-RTS and ARTS types of selection method depending on the knowledge of the turbulence statistics. The two selection approaches could be employed in conjunction depending on the knowledge of turbulence statistics. To maximize the extracted secure key rate, a conservative transmittance threshold might be used to perform P-RTS-type real time data rejection, followed by an ARTS-type scan during post selection.

It should be mentioned that our security assumption \cite{lo2007} relies on the random phase of Alice and Bob's quantum signals, which can be achieved by using phase modulators on both sides \cite{zhao2007}. Our setup does not randomize the phase of the quantum signals. However, our detection statistics are the same as a phase randomized system because the coherence times of Alice's and Bob's lasers are significantly shorter than the data collection time.

Altogether, overcoming atmospheric turbulence is crucial for establishing a global quantum network and the results presented here further demonstrate the capabilities of QKD systems in harsh environments.

\acknowledgements

We thank Eleftherios Moschandreou and Bing Qi for valuable discussions. This material is based upon work supported by the U.S. Department of Energy, Office of Science, Office of Advanced Scientific Computing Research, through the Quantum Internet to Accelerate Scientific Discovery Program under Field Work Proposal 3ERKJ381. We also acknowledge support by the National Science Foundation under award DGE-2152168.

\providecommand{\href}[2]{#2}\begingroup\raggedright\endgroup


\appendix


\section{Noise Model For Asymmetric Measurement Device Independent (MDI) QKD And Finite Key Calculation} \label{sec:A1}

In this Appendix, we establish a noise model to connect the 13 parameters for secret key calculation with a few QKD system parameters which can be easily calibrated. The QKD system parameters considered here are detector dark count rate / detection efficiency, polarization misalignment in the Z-basis, the HOM visibility in the X-basis. Note that HOM visibility depends on both the polarization misalignment in the X-basis, and the distinguishability of the photons from Alice and Bob. We denote the detector dark count rate as $Y_{0}$, detector efficiency as $\eta_{d}$ , polarization misalignment in the Z-basis (the probability that a H photon goes to V detector, or vice versa) as $e_{dz}$, and HOM visibility in the X-basis as $V_{HOM}$. To quantify the polarization misalignment, we use Charlie's polarization frame as a reference.

\subsection{Z-Basis}

When Alice and Bob send opposite polarization to Charlie the number of effective detection events:
\begin{equation*}
n_{z1} =   \frac{1}{2} N P_{s_{A}}P_{s_{B}} \Big(1-e^{-\eta_{A}\eta_{d}s_{A}} \Big) \Big(1-e^{-\eta_{B}\eta_{d}s_{B}}\Big)\Big(1 - 2e_{dz}\Big) 
\end{equation*}
The number of effective detection events when both Alice and Bob send the same polarization :
\begin{align*}
    n_{z2} =& \frac{1}{2} N P_{s_{A}}P_{s_{B}}\Big[1-e^{-(1-e_{dz})\eta_{A}\eta_{d}s_{A}} \; e^{-(1-e_{dz})\eta_{B}\eta_{d}s_{B}} \Big]  \\
    & \times\Big( e_{dz} \, \eta_{A} \, \eta_{d} \, s_{A}  +  Y_{0}  +  e_{dz} \, \eta_{B} \, \eta_{d} \, s_{B}  +  Y_{0} \Big)
\end{align*}
We denote the total detection count as $n_{ss}^{Z}$ and the error count as $m_{ss}^{Z}$ in Z basis. Here $n_{ss}^{Z} = n_{z1} + n_{z2}$ and $m_{ss}^{Z} = n_{z2}$.


\subsection{X-Basis}

We denote by $n_{i,j}^{X}$ the total number of detections given that Alice prepared $i$ photon state in $X$ basis and Bob prepared $j$ photon state in $X$ basis ($i,j = \mu, \nu, \omega$). In the X-basis, the correct detection for \{A, D\} and \{D, A\} is $\ket{\Psi^{-}}$, while the correct detection for \{A, A\} and \{D, D\} is $\ket{\Psi^{+}}$. When Alice and Bob prepare \{A,D\} or \{D,A\} we consider 3 cases based on the photon number arrived at the detectors.

\begin{enumerate}

\item  \textbf{\{1, 0\} and \{0, 1\} case:} \\
Corresponding probability of $\ket{\Psi^{-}}$ event:
\begin{equation*} \ Y_{0}  ( \eta_{d}  \eta_{A}  \mu_{A}  +  \eta_{d}  \eta_{B}  \omega_{B} )  e^{-\eta_{A} \eta_{d} \mu_{A} -  \eta_{B} \eta_{d} \omega_{B}}  
\end{equation*}
Corresponding probability of $\ket{\Psi^{+}}$ event:
\begin{equation*} 
  \ Y_{0} ( \eta_{d}  \eta_{A}  \mu_{A}  +  \eta_{d}  \eta_{B}  \omega_{B} )  e^{-\eta_{A}\eta_{d}\mu_{A} - \eta_{B}\eta_{d}\omega_{B} } 
\end{equation*}

\item \textbf{\{1, 1\} case:} \\
Corresponding probability of $\ket{\Psi^{-}}$ event: 
\begin{equation*}
\frac{1}{2}  (1-2e_{dX} )  \eta_{A} \eta_{d}\mu_{A}  \eta_{B} \eta_{d} \omega_{B}  e^{-\eta_{A} \eta_{d} \mu_{A}  -  \eta_{B} \eta_{d} \omega_{B} }
\end{equation*}
Corresponding probability of $\ket{\Psi^{+}}$ event:
\begin{equation*}
\frac{1}{4}   ( 1 - 2e_{dX} )  \eta_{d} \eta_{A} \mu_{A}   \eta_{d}\eta_{B}\mu_{B}  e^{-\mu_{A}\eta_{d}\mu_{A}-\eta_{B}\eta_{d}(\mu_{A} + \omega_{B})} 
\end{equation*}

\item \textbf{\{2, 0\} and \{0, 2\} case:} \\
Corresponding probability of $\ket{\Psi^{-}}$ event:
\begin{equation*}
\frac{1}{4}  \left[ (\eta_{d}  \eta_{A}  \mu_{A})^{2}  +  (\eta_{d}\eta_{B}\omega_{B})^{2} \right]  e^{- \mu_{A}  \eta_{d}  \mu_{A}  - \eta_{B}  \eta_{d} \omega_{B} }  
\end{equation*}
\noindent Corresponding probability of $\ket{\Psi^{+}}$ event:
\begin{equation*}
 \frac{1}{4}  e_{dX} \left[ ( \eta_{d} \eta_{A} \mu_{A} \big)^{2}  + \big( \eta_{d} \eta_{B} \omega_{B} \big)^{2} \right]  e^{-\mu_{A}\eta_{d}\mu_{A} -  \eta_{B}\eta_{d}\omega_{B}}   
\end{equation*}
   
\end{enumerate}

\begin{widetext}
Combining the above 3 cases, we obtain the total number of $\ket{\Psi^{-}}$ events:
\begin{align*}    
n_{c_{1}} =  & \frac{1}{2} N  P_{\mu_{A}}  P_{\omega_{B}} \Big[ Y_{0}  ( \eta_{d}  \eta_{A}  \mu_{A}  +  \eta_{d}  \eta_{B}  \omega_{B})  e^{-\eta_{A}\eta_{d}\mu_{A}  -  \eta_{B} \eta_{d} \omega_{B} }      + \frac{1}{2} ( 1-2e_{dX} )   \eta_{A} \eta_{d} \,\mu_{A}  \eta_{B} \eta_{d} \omega_{B}   e^{-\eta_{A} \eta_{d} \mu_{A}  - \eta_{B} \eta_{d} \omega_{B} }   \\
&  + \frac{1}{4} \left[ (\eta_{d} \eta_{A} \mu_{A})^{2} + (\eta_{d}\eta_{B}\omega_{B} )^{2} \right]  e^{-\mu_{A} \eta_{d} \mu_{A}  -  \eta_{B} \eta_{d} \omega_{B}}  \Big]
\end{align*}

\noindent and the total number of $\ket{\Psi^{+}}$ events: 
\begin{align*}
n_{w_{1}} = & \frac{1}{2} N P_{\mu_{A}} P_{\omega_{B}} \Big[ Y_{0} ( \eta_{d} \eta_{A} \mu_{A}  +  \eta_{d} \eta_{B} \omega_{B} )  e^{- \eta_{A} \eta_{d} \mu_{A}  -  \eta_{B} \eta_{d} \omega_{B} }    + \frac{1}{4} ( 1-2e_{dX} )  \eta_{d} \eta_{A} \mu_{A}   \eta_{d} \eta_{B} \mu_{B}  e^{-\mu_{A} \eta_{d} \mu_{A} -\eta_{B}\eta_{d}(\mu_{A}+\omega_{B})}         \\
& +\frac{1}{4} e_{dX} \left[ (\eta_{d} \eta_{A} \mu_{A} )^{2} +(\eta_{d} \eta_{B} \omega_{B} )^{2} \right] e^{-\mu_{A}\eta_{d}\mu_{A}-\eta_{B}\eta_{d}\omega_{B}} \Big]   
\end{align*}
When Alice and Bob prepare \{D, D\} or \{A, A\}, the analysis will be similar to \{A, D\} or \{D, A\}, except the roles of $\ket{\Psi^{+}}$ and $\ket{\Psi^{-}}$ are interchanged. Here $n_{c_{2}} = n_{c_{1}}$ and $n_{w_{2}} = n_{w_{1}}$. Therefore, $n_{i,j}^{X} = 2(n_{c_{1}}+n_{w_{1}})$ and $m_{i,j}^{X}=2n_{w_{1}}$. 

Combining the above all we obtain the total number of detection count:  
\begin{align*}
n_{\mu\omega}^{X} = & N P_{\mu_{A}} P_{\omega_{B}} \Big[ 2 Y_{0} (\eta_{d}  \eta_{A}  \mu_{A} + \eta_{d}  \eta_{B}  \omega_{B})  e^{-\eta_{A}\eta_{d}\mu_{A}- \eta_{B}\eta_{d}\omega_{B}}    +\frac{1}{4}(\eta_{d}\eta_{A}\mu_{A}+\eta_{d}\eta_{B}\omega_{B})^{2}e^{-\eta_{A}\eta_{d}\mu_{A}-\eta_{B}\eta_{d}\omega_{B}} \Big]
\end{align*}
Combining the above all we obtain the total number of error count:
\begin{align*}
m_{\mu\omega}^{X} = & N P_{\mu_{A}} P_{\omega_{B}}\Big[ Y_{0} ( \eta_{d}  \eta_{A}  \mu_{A}  +  \eta_{d}  \eta_{B}  \omega_{B} )  e^{-\eta_{A}\eta_{d}\mu_{A} - \eta_{B}\eta_{d}\omega_{B}}   \\
&  +\dfrac{1}{8} \left\{ (\eta_{d}\eta_{A}\mu_{A} )^{2} + ( \eta_{d}\eta_{B}\omega_{B} )^{2} + 8e_{dX}  \eta_{d}\eta_{A}\mu_{A}  \eta_{d} \eta_{B} \omega_{B}  \right\}    e^{-\eta_{A}\eta_{d}\mu_{A} - \eta_{B}\eta_{d}\omega_{B}} \Big]
\end{align*}
All the other terms can be determined by simply replacing all possible average photon numbers with the corresponding combinations of ($\mu, \nu, \omega$).
\end{widetext}

\begin{figure}[ht!]
    \centering
    \includegraphics[width=0.48\textwidth]{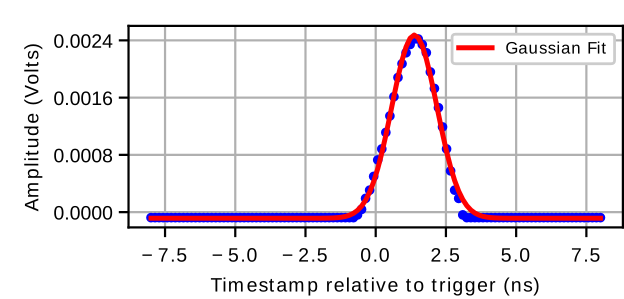}
    \includegraphics[width=0.48\textwidth]{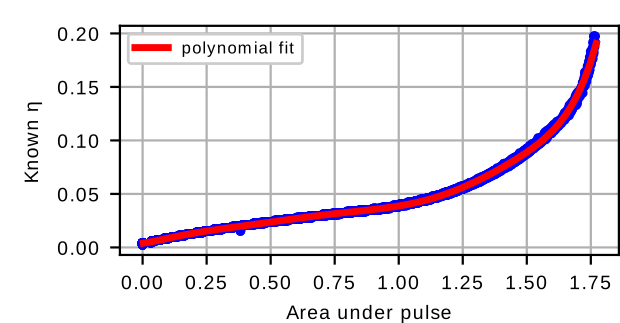}
    \caption{ Using classical probe pulses to estimate the quantum channel's transmittance. Upper panel: An example of Gaussian fit of a classical probe pulse obtained using the oscilloscope’s Fast-Frame feature. Lower panel: correlation between the programmed transmittance and the area under the pulse using a polynomial fit.  }
    \label{fig:4}
\end{figure}

\subsection{Finite Key Calculation}

Next, we account for the finite size effect using standard error analysis \cite{xu2013}. We denote the observed total counts and error counts as $n_{i,j}^{X}$ and $m_{i,j}^{X}$, respectively, where $i, j = \mu, \nu, \omega$.

The corresponding gains are
\begin{equation*}
Q_{i,j}^{X} =\frac{n_{i,j}^{X}}{NP_{i_{A}}P_{j_{B}}} \end{equation*}
and the errors are given by
\begin{equation*}
E_{i,j}^{X} =\frac{T_{i,j}^{X}}{Q_{i,j}^{X}} \ , \ \ 
T_{i,j}^{X}=\frac{m_{i,j}^{X}}{NP_{i_{A}}P_{j_{B}}}
\end{equation*}
The upper and lower bounds of gains  and errors are:
\begin{align*}
\overline{Q_{i,j}^{X}} &=Q_{i,j}^{X}+\gamma\sqrt{\frac{Q_{i,j}^{X}}{NP_{i_{A}}P_{j_{B}}}} \\
\underline{Q_{i,j}^{X}} &= Q_{i,j}^{X}-\gamma\sqrt{\frac{Q_{i,j}^{X}}{NP_{i_{A}}P_{j_{B}}}} \\
\overline{T_{i,j}^{X}} &= T_{i,j}^{X}+\gamma\sqrt{\frac{T_{i,j}^{X}}{NP_{i_{A}}P_{j_{B}}}} \\
\underline{T_{i,j}^{X}} &=T_{i,j}^{X}-\gamma\sqrt{\frac{T_{i,j}^{X}}{NP_{i_{A}}P_{j_{B}}}} 
\end{align*}
where $\gamma$ is related to the failure probability $\epsilon$ via $\epsilon = \text{erfc}(\frac{\gamma}{\sqrt{2}})$. We chose $\gamma=5.3$, so that $\epsilon \lesssim 10^{-7}$. 

The lower bound of yield $Y_{11}^{X,L}$ is estimated as follows:
\begin{equation*}
Y_{11}^{X,L}=\dfrac{1}{\mu - \nu}  \Big[ \dfrac{\mu+\omega}{(\nu-\omega)^{2}}  \underline{Q_{\nu\nu}^{M1}} + \frac{\nu+\omega}{(\mu - \omega)^{2}} \overline{Q_{\mu\mu}^{M2}} \Big] 
\end{equation*}
where
\begin{align*}
\underline{Q_{\nu\nu}^{M1}} &= e^{2\nu} \underline{Q_{\nu\nu}^{X}}  +  
e^{2\omega}  \underline{Q_{\omega\omega}^{X}} +  e^{\nu+\omega}  \overline{Q_{\nu\omega}^{X}} + e^{\omega+\nu}  \overline{Q_{\omega\nu}^{X}} \\
\overline{Q_{\mu\mu}^{M2}} &= e^{2\mu}  \overline{Q_{\mu\mu}^{X}}  +  e^{2\omega}  \underline{Q_{\omega\omega}^{X}}  -  e^{\mu+\omega}  \underline{Q_{\mu\omega}^{X}}  -  e^{\omega+\mu}  \underline{Q_{\omega\mu}^{X}} 
\end{align*}
The upper bound of error $e_{11}^{X,U}$ is estimated as:
\begin{equation*}
e_{11}^{X,U} = \frac{e^{2\nu} \, \overline{T_{\nu\nu}^{X}}  +  e^{2\omega}  \overline{ T_{\omega\omega}^{X}}  -  e^{\nu+\omega}  \underline{T_{\nu\omega}^{X}} - e^{\omega+\nu} \underline{T_{\omega\nu}^{X}}}{(\mu-\nu)Y_{11}^{X,L}} 
\end{equation*}
The final secure key rate is: 
\begin{align}
 R = P_{s_{A}}  P_{s_{B}} & \Big[ Y_{11}^{X,L} s^2 e^{-2s}  \big( 1-h_{2} (e_{11}^{X,U}) \big) \nonumber\\
 &\, - f \, Q_{ss}^{Z} \, h_{2}(E_{ss}^{Z}) \Big]  \label{eq:A18}
\end{align}

\section{Classical Channel}\label{sec:A2}

To estimate the channel's transmittance with classical probe pulses we follow the same recipe of \cite{lefty2021}. In our experimental setup, the classical probe pulses were set at a repetition rate of 4 kHz and a FWHM of $\sim$ 3 ns. The classical pulses were sent along with quantum pulses to the AOM using lTU channel 29. After reaching the AOM, the classical pulses were separated from the quantum pulses using a DWDM. The classical pulses were then sent to the high-gain classical photo-detector, which was connected to the DPO 7205 Tektronix Oscilloscope. The oscilloscope has a Fast-Frame feature to store high-resolution pulse data in a short (16 ns) interval around the trigger, sampled at 5 Giga samples per second (Fig.\ \ref{fig:4} (upper panel)). A Gaussian fit was performed on the measured classical pulses in order to calculate the area under the pulse to quantify the intensity. Finally, the transmittance was extracted from the measured pulse area using a polynomial fit. Fig.\ \ref{fig:4} (lower panel) shows that a similar resolution can be achieved by summing all the samples of each frame, with a significantly faster computation time than the Gaussian fit procedure.

\end{document}